\documentclass[A4paper,10pt,single spaced]{article}
\usepackage{amsmath,amssymb,amsfonts,amsthm,fancyhdr}
\usepackage{graphicx}
\def\ra{\rangle}
\def\la{\langle}
\def\be{\begin{equation}}
\def\ee{\end{equation}}
\def\ba{\begin{array}}
\def\ea{\end{array}}

\input amssym.def
\def\ra{\rangle}
\def\la{\langle}
\begin{document}
\begin{center}
{\Large \bf Characterization of four-qubit states via Bell inequalities }
\end{center}

\begin{center}
{\normalsize \author\small  Hui Zhao$^{1}$\ \  Xing-Hua Zhang$^{1}$\
\  Shao-Ming Fei$^{2}$
\ \   Zhi-Xi Wang$^{2}$}\\
 {\sl\small $^{1}$ College of Applied Sciences, Beijing University of Technology,
Beijing 100124, China\\
$^{2}$ School of Mathematical Sciences, Capital Normal University,
Beijing
100048, China }\\
\end{center}

\noindent{\bf Abstract} A set of Bell inequalities classifying the
quantum entanglement of four-qubit states is presented. These
inequalities involve only two measurement settings per observer and
can characterize fully separable, bi-separable and tri-separable
quantum states. In addition, a quadratic inequality of the Bell
operators for four-qubit systems is derived.


{PACS Number(s)}: 03.67.Mn, 03.65.Ud.

\section {Introduction}

The Bell inequality \cite{Bell64} provided the first possibility to
distinguish experimentally between quantum-mechanical predictions
and those of local realistic models. Derivations of new and stronger
Bell inequalities are one of the most important and challenging
subjects in quantum information processing. Since Bell's work, there
were many important generalizations such as \cite{MABK}-\cite{sci13}
and references therein.

The Bell inequalities presented in \cite{Chen2006} involve only two measurement settings
per observer and can detect perfectly the quantum entanglement
of the generalized GHZ states.
By using the idea in constructing Bell operators \cite{Chen2006},
a set of new Bell inequalities are given in \cite{lkch00}, which
gives rise to a finer classification of the entanglement for three-qubit systems.

The entanglement of four-qubit systems has been treated in terms of
Bell inequalities of Mermin-Klyshko type. In Ref. \cite{cfwu} the
quantum nonlocality of some four-qubit states, the GHZ state, W
state, cluster state and the state proposed in Ref. \cite{yyeo}, has
been investigated, towards the optimal violations of the Bell
inequality for these states. The classification of entanglement has
been also studied in such as Ref. \cite{yxxia}-\cite{sci21} with
linear inequalities for qubit systems and \cite{nha} with non-linear
inequalities for detecting bi-separable states in arbitrary
dimensional quantum systems.

In this paper, we study the quantum entanglement of four-qubit
systems by using the idea in constructing Bell operators in
\cite{Chen2006}. We generalize the results of three-qubit systems in
\cite{lkch00} to four-qubit systems. It has been shown that the
standard Werner-Wolf-$\dot{Z}$ukowski-Brukner (WWZB) inequalities
can not detect the entanglement of the generalized
Greenberger-Horne-Zeilinger (GHZ) states given by $|\psi \rangle
=\cos \alpha |0,...,0\rangle +\sin \alpha |1,...,1\rangle$ with
$0\!\leq \!\alpha \!\leq \!\pi /4$ \cite{SG2001,ZBLW2002}. However,
the Bell operators constructed in the way provided in
\cite{Chen2006} can detect the entanglement of the generalized GHZ
state wholly. Our Bell operators are constructed by using the idea
in \cite{Chen2006}. The resulted Bell inequalities can distinguish
fully separable, bi-separable and tri-separable states of a
four-qubit system. Moreover, these linear Bell inequalities involve
only two measurement settings per observer. Analytical formulas of
the average values of the Bell operators for four-qubit systems are
also derived. And a quadratic inequality of the Bell operators for
all four-qubit systems has been presented. Explicit geometrical
pictures show the relations between the different types of quantum
entanglement and the violations of the inequalities.

We fix some notations used in this paper. We use the Dirac's symbols
throughout this paper. If a quantum system is in one of a number of
states $|\psi\ra_i$, where $i$ is an index, with respective
probabilities $p_i$, then $\{p_i, |\psi\ra_i\}$ is called an
ensemble of pure states, and the associated density operator for the
system is defined by $\rho=\sum_i p_i \rho_i$. The average value of
the observable $M$ is written $\la M\ra=\la\psi|M|\psi\ra$.

\section {Classification of four qubits with Bell inequalities}

Consider $N$ parties and allow each of them to choose independently
between two dichotomic observables $A_{j}$, $B_{j}$ for the $j$-th
observer, where $A_j=\vec{a}_j\cdot\vec{\sigma}^j$ and
$B_j=\vec{b}_j\cdot\vec{\sigma}^j$, with
$\vec{\sigma}^j=({\sigma_1}^{(j)},{\sigma_2}^{(j)},{\sigma_3}^{(j)})$
the Pauli matrices on the $j$-th qubit, and
$\vec{a}_j=(a^{(1)}_j,a^{(2)}_j,a^{(3)}_j)$,
$\vec{b}_j=(b^{(1)}_j,b^{(2)}_j,b^{(3)}_j)$ the real unit vectors.
The quantum mechanical Bell operator on the $N-1$ qubits except for
the $i$-th qubit is defined as ${\cite{Chen2006}}$
$$
D_{N}^{(i)}=B_{N-1}^{(i)}\otimes\frac{1}{2}(A_i+B_i)+I_{N-1}\otimes\frac{1}{2}(A_i-B_i),\quad
i=1,\ldots, N,
$$
where $B_{N-1}^{(i)}$ is the Bell operator of WWZB inequalities on
the $N-1$ qubits except for the $i$-th qubit,
$$\begin{array}{ll}
B_{N-1}^{(i)} =&\displaystyle\frac{1}{2^{N-1}}\sum_{S_1,\ldots, S_{N-1}={-1,1}}
S(S_1,\ldots,S_{N-1})\\[6mm]
&\displaystyle\sum_{K_1,\ldots, K_{N-1}={1,2}}{S_1}^{{K_1}-1} \ldots
{S_{N-1}}^{K_{N-1}-1}\otimes_{j=1}^{N-1}O_j(K_j),
\end{array}
$$
where $I_{N-1}$ denotes the corresponding identity matrix. For
$S(S_1,\ldots,
S_{N-1})=\sqrt{2}\cos(\frac{\pi}{4}(S_1+\cdots+S_N-N)-\frac{\pi}{4})$,
$O_j(1)=A_j$, $O_j(2)=B_j$, one recovers the
Mermin-Ardehali-Belinskii-Klyshko (MABK) inequalities \cite{MABK}.

In the following we study the
characterization of entanglement for four-qubit systems, $N=4$.

\noindent{{\bf Theorem 1.}} For fully separable states $\rho$, we
have
\be\label{t1}
|\langle{D_4}^{(i)}\rangle|\leq1,\ \ i=1,2,3,4.
\ee

\noindent{{{\bf Proof.}}} A general pure four-qubit state can be
written as $|\psi\ra=\sum_{i,j,k,l=0}^1 a_{ijkl}|ijkl\ra$ with
normalization $\sum_{i,j,k,l=0}^1 |a_{ijkl}|^2=1$. A mixed
four-qubit state can be expressed as $\rho=\sum_\alpha p_\alpha
\rho_\alpha$, where $0<p_\alpha\leq 1$, $\sum_\alpha p_\alpha=1$,
$\rho_\alpha=|\psi_\alpha\ra \la\psi_\alpha|$ are pure states. Due
to the linear property of the average values,
$$
\begin{array}{ll}
|\langle{D_4}^{(i)}\rangle| &=|tr(\sum p_\alpha\rho_\alpha D_4^{(i)})|=|\sum p_\alpha tr(\rho_ \alpha D_4^{(i)})|\\[3mm]
&\leq\sum|p_\alpha tr(\rho_\alpha D_4^{(i)})|\leq\sum|tr(\rho_\alpha
D_4^{(i)})|,
\end{array}
$$
it is sufficient to consider the pure states. The fully separable
pure states can be transformed into the form $|\psi\ra=|0000\ra$ in
suitable bases. Therefore it is direct to verify that
$$
\begin{array}{ll}
 |{\langle{D_4}^{(1)}\rangle}|
 =&|\frac{1}{4}((a_1^{(3)}+b_1^{(3)})(-a_2^{(3)}a_3^{(3)}a_4^{(3)}+a_2^{(3)}b_3^{(3)}b_4^{(3)}\\[3mm]
&+b_2^{(3)}a_3^{(3)}b_4^{(3)}+b_2^{(3)}b_3^{(3)}a_4^{(3)})+\frac{1}{2}(a_1^{(3)}-b_1^{(3)})|
\leq 1.
\end{array}
$$
Similarly one can prove that $|{\langle{D_4}^{(i)}\rangle}|\leq 1$
for $i=2,3,4.$ $\quad \Box$

Next we consider the cases of tri-separable states. We denote
$\rho_{ij-k-l}$ a tri-separable state of the form $\rho_{ij}\otimes
\rho_k\otimes \rho_l$, in which qubits $i$ and $j$ are entangled,
while qubits $k$ and $l$ are separable, $i\neq j\neq k\neq
l=1,2,3,4$.

\noindent{\bf Theorem 2.} For any tri-separable states
$\rho_{ij-k-l}$, $i\neq j\neq k\neq l=1,2,3,4$, we have
\be\label{t2}
|\langle{D_4}^{(i)}\rangle|=|\langle{D_4}^{(j)}\rangle|\leq 1,~~
|\langle{D_4}^{(k)}\rangle|=|\langle{D_4}^{(l)}\rangle|\leq
\frac{3}{2}. \ee

\noindent{\bf Proof.} We consider the case of $\rho_{12-3-4}$. Every
pure state in $\rho_{12-3-4}$ can be written in a Schmidt form,$
|\psi\ra=(\cos\alpha|01\ra-\sin\alpha|10\ra)\otimes|0\ra\otimes|0\ra.
$ Therefore
$$
\begin{array}{lll}
 && |{\langle{D_4}^{(1)}\rangle}_{|\psi\ra}|\\[3mm]
     &=&{|\frac{1}{2}}[-(a_1^{(3)}a_2^{(3)}+a_2^{(3)}b_1^{(3)})-(\sum_{k=1}^2a_1^{(k)}a_2^{(k)}
   +a_2^{(k)}b_1^{(k)})\sin{2\alpha}]\frac{1}{2}(-a_3^{(3)}a_4^{(3)}+b_3^{(3)}b_4^{(3)})\\[3mm]
&&
+\frac{1}{2}[-(a_1^{(3)}b_2^{(3)}+b_1^{(3)}b_2^{(3)})-(\sum_{k=1}^2a_1^{(k)}b_2^{(k)}
+b_1^{(k)}b_2^{(k)})\sin{2\alpha}]\frac{1}{2}(a_3^{(3)}b_4^{(3)}+b_3^{(3)}a_4^{(3)})\\[3mm]
 &&+\frac{1}{2}(a_1^{(3)}-b_1^{(3)})\cos{2\alpha}|\leq 1.
 \end{array}
$$
Similarly one can prove that
$|\langle{D_4}^{(2)}\rangle_{|\psi\ra}|\leq1$. For the Bell operator
$D_4^{(3)}$, we have
$$
\begin{array}{lll}
&&|{\langle{D_4}^{(3)}\rangle}_{|\psi\ra}|\\[3mm]
  &=&|{\frac{1}{4}}\{[-(-a_1^{(3)}a_2^{(3)}+b_1^{(3)}b_2^{(3)})
  -(\sum_{k=1}^2-a_1^{(k)}a_2^{(k)}+b_1^{(k)}b_2^{(k)})\sin{2\alpha}][(a_3^{(3)}+b_3^{(3)})a_4^{(3)}]\\[3mm]
&&+[ -(a_1^{(3)}b_2^{(3)}+b_1^{(3)}a_2^{(3)})
-(\sum_{k=1}^2a_1^{(k)}b_2^{(k)}+b_1^{(k)}a_2^{(k)})\sin{2\alpha}][(a_3^{(3)}+b_3^{(3)})b_4^{(3)}]\}\\[3mm]
&&+\frac{1}{2}(a_3^{(3)}-b_3^{(3)})|\leq \frac{3}{2}.
\end{array}$$
In a similar way, we have
$|\langle{D_4}^{(4)}\rangle_{|\psi\ra}|\leq\frac{3}{2}$. The cases
of $\rho_{1-2-34}$, $\rho_{13-2-4}$, $\rho_{1-3-24}$,
$\rho_{14-2-3}$ and $\rho_{1-4-23}$ can be similarly proved.
According to the linear property of average values, for all
tri-separable states we have (\ref{t2}). \quad $\Box$

Finally we consider the cases of bi-separable states. There are two classes of bi-separable ones.
i) Two entangled qubits $i$ and $j$ are separable from other entangled qubits $k$ and $l$.
For example, we denote $\rho_{12-34}$ the bi-separable state of
the form $\rho_{12}\otimes \rho_{34}$, where the qubits $12$ and $34$ are entangled
respectively.
ii) A qubit $i$ is separable from the rest genuine tripartite entangled qubits $j$, $k$ and $l$.
For instance, $\rho_{1-234}$ denotes a bi-separable state of the
form $\rho_{1}\otimes \rho_{234}$, where qubits $234$ are genuine entangled.

\noindent{\bf Theorem 3.} For all bi-separable states $\rho$, we
have \be\label{t31} |{\langle{D_4}^{(i)}\rangle}_{\rho}|\leq
\frac{3}{2},\quad i=1,2,3,4, \ee for $\rho$ in class i), and
\be\label{t32}
|{\langle{D_4}^{(i)}\rangle}_{\rho}|\leq\sqrt{3},\quad i=1,2,3,4,
\ee for $\rho$ in class ii).

\noindent{\bf Proof.} We first consider bi-separable states in class
i). A pure state in $\rho_{12-34}$ has the following general form, $
|\psi\ra=(\cos\alpha|01\ra-\sin\alpha|10\ra)\otimes(\cos\beta|01\ra-\sin\beta|10\ra).
$ Hence
$$
\begin{array}{lll}
  |{\langle{D_4}^{(1)}\rangle}_{|\psi\ra}|
&=&|{\frac{1}{4}}\{[-({{a_1}^{(3)}}{{a_2}^{(3)}}+{{a_2}^{(3)}}{{b_1}^{(3)}})
-(\sum_{k=1}^2{{a_1}^{(k)}}{{a_2}^
{(k)}}+{a_2}^{(k)}{b_1}^{(k)}){\sin{2\alpha}}]\\[3mm]
& &\times[-({{-a_3}^{(3)}}{{a_4}^{(3)}}+{{b_3}^{(3)}}{{b_4}^{(3)}})
-(\sum_{k=1}^2-{{a_3}^{(k)}}{{a_4}^
{(k)}}+{b_3}^{(k)}{b_4}^{(k)}){\sin{2\beta}}]\\[3mm]
& &+[-({{a_1}^{(3)}}{{b_2}^{(3)}}+{{b_1}^{(3)}}{{b_2}^{(3)}})
-(\sum_{k=1}^2{{a_1}^{(k)}}{{b_2}^
{(k)}}+{b_1}^{(k)}{b_2}^{(k)}){\sin{2\alpha}}]\\[3mm]
& &\times[-({{a_3}^{(3)}}{{b_4}^{(3)}}+{{a_4}^{(3)}}{{b_3}^{(3)}})
-(\sum_{k=1}^2{{a_3}^{(k)}}{{b_4}^
{(k)}}+{a_4}^{(k)}{b_3}^{(k)}){\sin{2\beta}}]\}\\[3mm]
& &+\frac{1}{2}{({a_1}^{(3)}-{{b_1}^{(3)}}){\cos{2\alpha}}}|\leq
\frac{3}{2}.
\end{array}
$$
Similarly one can get
$|\langle{D_4}^{(i)}\rangle_{|\psi\ra}|\leq\frac{3}{2}$ for
$i=2,3,4$. For the cases of $\rho_{13-24}$ and $\rho_{14-23}$, we
can also similarly have $|\langle{D_4}^{(i)}\rangle_{|\psi\ra}|\leq
\frac{3}{2}$ for $i=1,2,3,4.$

For bi-separable states in class ii), we consider the case of $\rho_{1-234}$.
There are two inequivalent kinds of genuine three-qubit
entangled states, the GHZ-type and W-type $\cite{wdur}$. For simplicity in the following
we denote $c_x=\cos x$ and $s_x=\sin x$.
The GHZ-type state can be written as
$$
|\psi_{GHZ}\rangle=\sqrt{K}(c_\delta
|0\rangle|0\rangle|0\rangle+s_\delta
e^{i\varphi}|\phi_A\rangle|\phi_B\rangle|\phi_C\rangle),
$$
where $|\phi_A\rangle=c_\alpha|0\rangle+s_\alpha|1\rangle,~~~
|\phi_B\rangle=c_\beta|0\rangle+s_\beta|1\rangle,~~~
|\phi_C\rangle=c_\gamma|0\rangle+s_\gamma|1\rangle,$
$\delta\in(0,\frac{\pi}{4}]$,
$\alpha,\beta,\gamma\in(0,\frac{\pi}{2}]$, $\varphi\in[0,2\pi)$ and
$K=(1+2c_\delta s_\delta c_\alpha c_\beta c_\gamma
c_\varphi)^{-1}\in (\frac{1}{2},\infty)$ is a normalization factor.
The W-type state can be written as
$$
|\psi_{W}\rangle=\sqrt{a}|001\rangle+\sqrt{b}|010\rangle
+\sqrt{c}|100\rangle+\sqrt{d}|000\rangle,
$$
where $a, b, c>0$ and $d=1-(a+b+c)\geq 0.$ Therefore every pure
state in $\rho_{1-234}$ via a suitable choice of bases can be
written as $|\psi_{0-GHZ}\rangle=|0\ra\otimes|\psi_{GHZ}\rangle,~~~
or~~~ |\psi_{0-W}\rangle=|0\ra\otimes|\psi_{W}\rangle.$

We calculate here the value $|\la A_{2}A_3A_4\ra_{|\psi_{GHZ}\ra}|$.
The other items have similar expressions.
$$
\begin{array}{l}
|\la A_2A_3A_4\ra_{|\psi_{GHZ}\ra}|=\\[3mm]
K\left |a_2^{(3)}a_3^{(3)}a_4^{(3)}(c_{\delta}^2+s_{\delta}^2\cdot\cos2\alpha\cdot\cos2\beta\cdot\cos2\gamma)\right.\\[3mm]
+a_2^{(1)}a_3^{(3)}a_4^{(3)}(\sin2\delta\cdot c_{\varphi}s_{\alpha}c_{\beta}c_{\gamma}+s_{\delta}^2\cdot\sin2\alpha\cdot\cos2\beta\cdot\cos2\gamma)\\[3mm]
+a_2^{(3)}a_3^{(1)}a_4^{(3)}(\sin2\delta\cdot c_{\varphi}c_{\alpha}s_{\beta}c_{\gamma}+s_{\delta}^2\cdot\cos2\alpha\cdot\sin2\beta\cdot\cos2\gamma)\\[3mm]
+a_2^{(3)}a_3^{(3)}a_4^{(1)}(\sin2\delta\cdot c_{\varphi}c_{\alpha}c_{\beta}s_{\gamma}+s_{\delta}^2\cdot\cos2\alpha\cdot\cos2\beta\cdot\sin2\gamma)\\[3mm]
+a_2^{(1)}a_3^{(1)}a_4^{(3)}(\sin2\delta\cdot
c_{\varphi}s_{\alpha}s_{\beta}s_{\gamma}
+s_{\delta}^2\cdot\sin2\alpha\cdot\sin2\beta\cdot\cos2\gamma)
-a_2^{(2)}a_3^{(2)}a_4^{(3)}\sin2\delta\cdot c_{\varphi}s_{\alpha}s_{\beta}c_{\gamma}\\[3mm]
+a_2^{(1)}a_3^{(3)}a_4^{(1)}(\sin2\delta\cdot
c_{\varphi}s_{\alpha}c_{\beta}s_{\gamma}
+s_{\delta}^2\cdot\sin2\alpha\cdot\cos2\beta\cdot\sin2\gamma)
-a_2^{(2)}a_3^{(3)}a_4^{(2)}\sin2\delta\cdot c_{\varphi}s_{\alpha}c_{\beta}s_{\gamma}\\[3mm]
+a_2^{(3)}a_3^{(1)}a_4^{(1)}(\sin2\delta\cdot
c_{\varphi}c_{\alpha}s_{\beta}s_{\gamma}
+s_{\delta}^2\cdot\cos2\alpha\cdot\sin2\beta\cdot\sin2\gamma)
-a_2^{(3)}a_3^{(2)}a_4^{(2)}\sin2\delta\cdot c_{\varphi}c_{\alpha}s_{\beta}s_{\gamma}\\[3mm]
+a_2^{(1)}a_3^{(1)}a_4^{(1)}s_{\delta}^2\cdot\sin2\alpha\cdot\sin2\beta\cdot\sin2\gamma)\\[3mm]
+(a_2^{(1)}a_3^{(1)}a_4^{(1)}-a_2^{(2)}a_3^{(2)}a_4^{(1)}-a_2^{(2)}a_3^{(1)}a_4^{(2)}
\left. -a_2^{(1)}a_3^{(2)}a_4^{(2)})\sin2\delta\cdot
c_{\varphi}s_{\alpha}s_{\beta}s_{\gamma}\right|.
\end{array}
$$
$|\la A_2A_3A_4\ra_{|\psi_{GHZ}\ra}|$ attains its maximum at
$\alpha=\beta=\gamma=\frac{\pi}{2},$ $\varphi=0$ and
$\delta=\frac{\pi}{4}$ according to the value of $K$ and the
property of the trigonometric functions. Hence we have
$$
\begin{array}{lll}
  |{\langle{D_4}^{(1)}\rangle}_{|\psi_{0-GHZ}\rangle}|&\leq &
  |{\frac{1}{4}}(a_1^{(3)}+b_1^{(3)})[a_2^{(1)}(a_3^{(1)}a_4^{(1)}-a_3^{(2)}a_4^{(2)}-b_3^{(1)}b_4^{(1)}+
b_3^{(2)}b_4^{(2)})\\[3mm]&&+a_2^{(2)}(-a_3^{(1)}a_4^{(2)}-a_3^{(2)}a_4^{(1)}+
b_3^{(1)}b_4^{(2)}+b_3^{(2)}b_4^{(1)})\\[3mm]
&&+b_2^{(1)}(-a_3^{(1)}b_4^{(1)}+a_3^{(2)}b_4^{(2)}
-b_3^{(1)}a_4^{(1)}+b_3^{(2)}a_4^{(2)})\\[3mm]&&+b_2^{(2)}(a_3^{(1)}b_4^{(2)}+a_3^{(2)}b_4^{(1)}+
b_3^{(1)}a_4^{(2)}+b_3^{(2)}a_4^{(1)})]+\frac{1}{2}(a_1^{(3)}-b_1^{(3)})|\leq
1.
\end{array}%\eqno{(11)}
$$
Using the similar method above, we have also
$|\langle{D_4}^{(i)}\rangle_{|\psi_{0-GHZ}\rangle}|\leq1$ for
$i=2,3,4$.

Next we compute $|\la A_{2}A_3A_4\ra_{|\psi_W\ra}|$.
$$
\begin{array}{lll}
& &|\la A_2A_3A_4\ra_{|\psi_W\ra}|\\[3mm]
&=&|(d-a-b-c)a_2^{(3)}a_3^{(3)}a_4^{(3)}+2\sqrt{cd}a_2^{(1)}a_3^{(3)}a_4^{(3)}
+2\sqrt{bd}a_2^{(3)}a_3^{(1)}a_4^{(3)}\\[3mm]
&&+2\sqrt{ad}a_2^{(3)}a_3^{(3)}a_4^{(1)}+2\sqrt{bc}(a_2^{(1)}a_3^{(1)}a_4^{(3)}+a_2^{(2)}a_3^{(2)}a_4^{(3)})\\[3mm]
&&+2\sqrt{ac}(a_2^{(1)}a_3^{(3)}a_4^{(1)}+a_2^{(2)}a_3^{(3)}a_4^{(2)})
+2\sqrt{ab}(a_2^{(3)}a_3^{(1)}a_4^{(1)}+a_2^{(3)}a_3^{(2)}a_4^{(2)})|\\[3mm]
&\leq&
\frac{1}{2}|a_2^{(3)}(-a_3^{(3)}a_4^{(3)}+a_3^{(1)}a_4^{(1)}+a_3^{(2)}a_4^{(2)}+a_3^{(1)}a_4^{(3)}+a_3^{(3)}a_4^{(1)})\\[3mm]
&&+a_2^{(1)}(a_3^{(3)}a_4^{(3)}+a_3^{(1)}a_4^{(3)}+a_3^{(3)}a_4^{(1)})+a_2^{(2)}(a_3^{(2)}a_4^{(3)}+a_3^{(3)}a_4^{(2)})|\\[3mm]
&\leq&\frac{1}{2}|a_2^{(3)}+a_2^{(1)}+a_2^{(2)}|\leq
\frac{\sqrt{3}}{2},
\end{array}
$$
where we have used the Cauchy-Schwarz inequality and the relation
$\sqrt[4]{abcd}\leq \frac{a+b+c+d}{4}$, for which the equality holds
if and only if $a=b=c=d$. Therefore we have
$$
|{\langle{D_4}^{(1)}\rangle}_{|\psi_{0-W}\rangle}|\leq
  {\frac{1}{4}}\times2\times(\frac{\sqrt{3}}{2}\times4)=\sqrt{3}.$$
Similarly one can obtain
$|\langle{D_4}^{(i)}\rangle_{|\psi_{0-W}\rangle}|\leq\sqrt{3}$ for
$i=2,3,4$.

The cases for $\rho_{2-134},$ $\rho_{3-124}$ and $\rho_{4-123}$ can
be similarly proved. \quad $\Box$

\section {The quadratic inequality of Bell operator for four qubits}

We derive now an analytical quadratic inequality of the Bell operator
for four qubits. The four-qubit states $\rho$ can be written as \cite{linden},
\be\label{13}
\begin{array}{ll}
  \rho &= {\frac{1}{16}}(I\otimes{I}\otimes{I}\otimes{I}+\sum_{i_1=1}^{3}Q_{i_1}^{(1)}\sigma_{i_1}^{(1)}
  \otimes{I}\otimes{I}\otimes{I}\\[3mm]
  &+\sum_{i_2=1}^{3}Q_{i_2}^{(2)}I \otimes\sigma_{i_2}^{(2)}\otimes{I}\otimes{I}+\sum_{i_3=1}^{3}Q_{i_3}^{(3)}I
  \otimes{I}\otimes{\sigma_{i_3}^{(3)}}\otimes{I}+\sum_{i_4=1}^{4}Q_{i_4}^{(4)}I
  \otimes{I}\otimes I\otimes{{\sigma_{i_4}^{(4)}}}\\[3mm]
  &+\cdots+
\sum_{i_1,i_2,i_3,i_4=1}^{3}Q_{i_1i_2i_3i_4}^{(1234)}\sigma_{i_1}^{(1)}\otimes\sigma_{i_2}^{(2)}\otimes{\sigma_{i_3}
^{(3)}}\otimes\sigma_{i_4}^{(4)}).
\end{array}
\ee
Set
$$
\begin{array}{l}
  \overrightarrow{\alpha}=(Q_1^{(1)},Q_2^{(1)},Q_3^{(1)}),\ \ \overrightarrow{\beta}=(Q_1^{(2)},Q_2^{(2)},Q_3^{(2)}),\\[3mm]
\overrightarrow{\gamma}=(Q_1^{(3)},Q_2^{(3)},Q_3^{(3)}),\ \ \overrightarrow{\varepsilon}=(Q_1^{(4)},Q_2^{(4)},Q_3^{(4)}),\\[3mm]
\overrightarrow{S}=(Q_{111}^{(124)},Q_{112}^{(124)},Q_{113}^{(124)},\ldots,Q_{331}^{(124)},Q_{332}^{(124)},Q_{333}^{(124)}),\\[3mm]
\overrightarrow{T}=(Q_{111}^{(123)},Q_{112}^{(123)},Q_{113}^{(123)},\ldots,Q_{331}^{(123)},Q_{332}^{(123)},Q_{333}^{(123)}),\\[3mm]
\overrightarrow{U}=(Q_{111}^{(234)},Q_{112}^{(234)},Q_{113}^{(234)},\ldots,Q_{331}^{(234)},Q_{332}^{(234)},Q_{333}^{(234)}),\\[3mm]
\overrightarrow{V}=(Q_{111}^{(134)},Q_{112}^{(134)},Q_{113}^{(134)},\ldots,Q_{331}^{(134)},Q_{332}^{(134)},Q_{333}^{(134)}),\\[3mm]
\overrightarrow{Q}=(Q_{1111}^{(1234)},Q_{1112}^{(1234)},Q_{1113}^{(1234)},\ldots,\ldots,Q_{3331}^{(1234)},Q_{3332}^{(1234)},Q_{3333}^{(1234)}).
\end{array}
$$
We have the following lemma.

\noindent{\bf Lemma.} For four qubits pure states, we have the following equality,
\be\label{rl}
{{|{\overrightarrow{\alpha}}|}^2+{|{\overrightarrow{\beta}}|}^2
  +{|{\overrightarrow{\gamma}}|}^2+{|{\overrightarrow{\varepsilon
}}|}^2+{|{\overrightarrow{S}}|}^2+{|{\overrightarrow{T}}|}^2+{|{\overrightarrow{U}}|}^2
  +{|{\overrightarrow{V}}|}^2+|{\overrightarrow{Q}}|}^2=9.
\ee

\noindent{{{\bf Proof.}}} A four-qubit pure state $|\psi\rangle$
can be also written as {\cite{rudolph}},
$$
\begin{array}{ll}
|\psi\ra&=l_0{|0000\ra}+l_1{|0011\ra}+l_2{|0101\ra}+l_3{|0110\ra}+l_4{|1100\ra}+l_5{|1001\ra}+l_6{|1010\ra}\\[3mm]
&+l_7{|1011\ra}+l_8{|0111\ra}+l_9{|1101\ra}
+l_{10}{|1110\ra}+l_{11}{|1111\ra},
\end{array}
$$
where ${l_i}$ with $i=7,8,9,10,11$, are non-negative real numbers
and $ |l_0|\geq |l_i|$ for $i=1,2,\ldots,11$. Comparing
$\rho=|\psi\rangle\langle\psi|$ with (\ref{13}), we have the
relation (\ref{rl}) by straightforward calculation.

Here ${|{\overrightarrow{\alpha}}|}^2$,
${|{\overrightarrow{\beta}}|}^2$, ${|{\overrightarrow{\gamma}}|}^2$,
${|{\overrightarrow{\varepsilon }}|}^2$,
${|{\overrightarrow{S}}|}^2$, ${|{\overrightarrow{T}}|}^2$,
${|{\overrightarrow{U}}|}^2$, ${|{\overrightarrow{V}}|}^2$,
${|{\overrightarrow{Q}}|}^2$ are all invariants under local unitary
transformations, and equality (\ref{rl}) holds
for all pure states. The minimum of $|{\overrightarrow{Q}}|$ is
attained for fully separable states and the maximum of $|{\overrightarrow{Q}}|$
is obtained for maximally entangled states.

\noindent{\bf Theorem 4.} Any four-qubit mixed state $\rho$ satisfies
the following inequality,
\be\label{t4}
\omega={\langle{D_4}^{(1)}\rangle}_{\rho}^2+{\langle{D_4}^{(2)}\rangle}_{\rho}^2+{\langle{D_4}^{(3)}\rangle}_{\rho}^2+{\langle{D_4}^{(4)}\rangle}
 _{\rho}^2\leq4.
\ee

\noindent{\bf Proof.} Due to that the quadratic function is a convex
function,
$$
\begin{array}{ll}
{\langle{D_4}^{(i)}\rangle}^2 &={[tr(\sum p_\alpha\rho_\alpha D_4^{(i)})}]^2={[\sum p_\alpha tr(\rho_ \alpha D_4^{(i)})}]^2\\[3mm]
&\leq\sum p_\alpha {[tr(\rho_\alpha D_4^{(i)})]}^2\leq\sum
{[tr(\rho_\alpha D_4^{(i)})]}^2,
\end{array}
$$
it is sufficient to consider only pure states.
 Set
$$
C_i=\frac{1}{2}{(A_i+B_i)},\ \  D_i=\frac{1}{2}{(B_i-A_i)},\ \
\overrightarrow{s_i}=\frac{1}{2}{(\overrightarrow{b_i}+\overrightarrow{a_i})},\
\
\overrightarrow{t_i}=\frac{1}{2}{(\overrightarrow{b_i}-\overrightarrow{a_i})},
\eqno{(8)}$$ We have
${|{\overrightarrow{s_i}}|}^2+{|{\overrightarrow{t_i}}|}^2=1,$
${\overrightarrow{s_i}}\cdot{\overrightarrow{t_i}}=0$, and
$$
\begin{array}{ll}
{\langle{D_4}^{(1)}\rangle}_{\rho}^2
=&\{[\overrightarrow{s_1}\otimes\overrightarrow{s_2}\otimes
(\overrightarrow{s_3}\otimes\overrightarrow{t_4}-\overrightarrow{t_3}
\otimes\overrightarrow{t_4}+\overrightarrow{s_3}
\otimes\overrightarrow{s_4}+\overrightarrow{t_3}
\otimes\overrightarrow{s_4})\\[3mm]
&+\overrightarrow{s_1}\otimes\overrightarrow{t_2}\otimes(\overrightarrow{s_3}
\otimes\overrightarrow{s_4}-\overrightarrow{t_3}
\otimes\overrightarrow{s_4}-\overrightarrow{t_3}
\otimes\overrightarrow{t_4}-\overrightarrow{s_3}
\otimes\overrightarrow{t_4})]\cdot\overrightarrow{Q}-
\overrightarrow{t_1}\cdot\overrightarrow{\alpha}\}^2,
\end{array}$$
$$\begin{array}{ll}
{\langle{D_4}^{(2)}\rangle}_{\rho}^2
=&\{[\overrightarrow{t_1}\otimes\overrightarrow{s_2}\otimes(\overrightarrow{s_3}
\otimes\overrightarrow{s_4}-\overrightarrow{t_3}
\otimes\overrightarrow{s_4}-\overrightarrow{s_3}
\otimes\overrightarrow{t_4}-\overrightarrow{t_3}
\otimes\overrightarrow{t_4})\\[3mm]
&+\overrightarrow{s_1}\otimes\overrightarrow{s_2}\otimes(\overrightarrow{s_3}
\otimes\overrightarrow{t_4}-\otimes\overrightarrow{t_3}
\otimes\overrightarrow{t_4}+\overrightarrow{s_3}
\otimes\overrightarrow{s_4}+\overrightarrow{t_3}
\otimes\overrightarrow{s_4})]\cdot\overrightarrow{Q}-
\overrightarrow{t_2}\cdot\overrightarrow{\beta}\}^2,
\end{array}$$
$$\begin{array}{ll}
{\langle{D_4}^{(3)}\rangle}_{\rho}^2=&\{[\overrightarrow{t_1}\otimes\overrightarrow{s_2}\otimes(\overrightarrow{s_3}
\otimes\overrightarrow{s_4}-\overrightarrow{s_3}
\otimes\overrightarrow{t_4})+
\overrightarrow{s_1}\otimes\overrightarrow{t_2}\otimes(\overrightarrow{s_3}
\otimes\overrightarrow{s_4}-\overrightarrow{s_3}
\otimes\overrightarrow{t_4})\\[3mm]
&+
\overrightarrow{s_1}\otimes\overrightarrow{s_2}\otimes(\overrightarrow{s_3}
\otimes\overrightarrow{s_4}+\overrightarrow{s_3}
\otimes\overrightarrow{t_4})
-\overrightarrow{t_1}\otimes\overrightarrow{t_2}\otimes(\overrightarrow{s_3}
\otimes\overrightarrow{s_4}-\overrightarrow{s_3}
\otimes\overrightarrow{t_4})]\cdot\overrightarrow{Q}-
\overrightarrow{t_3}\cdot\overrightarrow{\gamma}\}^2,
\end{array}$$
$$\begin{array}{ll}
{\langle{D_4}^{(4)}\rangle}_{\rho}^2=&
\{[\overrightarrow{t_1}\otimes\overrightarrow{s_2}\otimes(\overrightarrow{s_3}
\otimes\overrightarrow{s_4}-\overrightarrow{t_3}
\otimes\overrightarrow{s_4})+
\overrightarrow{s_1}\otimes\overrightarrow{t_2}\otimes(\overrightarrow{s_3}
\otimes\overrightarrow{s_4}-\overrightarrow{t_3}
\otimes\overrightarrow{s_4})\\[3mm]
&+\overrightarrow{s_1}\otimes\overrightarrow{s_2}\otimes(\overrightarrow{s_3}
\otimes\overrightarrow{s_4}+\overrightarrow{t_3}
\otimes\overrightarrow{s_4})
-\overrightarrow{t_1}\otimes\overrightarrow{t_2}\otimes(\overrightarrow{s_3}
\otimes\overrightarrow{s_4}-\overrightarrow{t_3}
\otimes\overrightarrow{s_4})]\cdot\overrightarrow{Q}-
\overrightarrow{t_4}\cdot\overrightarrow{\varepsilon}\}^2,
\end{array}$$
where
$\overrightarrow{s}\otimes\overrightarrow{t}\otimes\overrightarrow{p}\otimes\overrightarrow{q}\cdot\overrightarrow{Q}$
denotes $\sum_{ijkh}s_{i}t_{j}p_{k}q_{h}Q_{ijkh}.$ $\omega$ attains
its maximum at either $|\overrightarrow{Q}|=1$ or
$|\overrightarrow{Q}|=3.$ For the case of $|\overrightarrow{Q}|=1,$
$|\psi\ra$ is fully separable and the inequality is satisfied by
using Theorem $1$. For the case of $|\overrightarrow{Q}|=3,$
$|\psi\ra$ is maximally entangled. Without losing generality, we
consider the maximally entangled state
$|\psi\ra=\frac{1}{\sqrt{2}}{(|0000\ra+|1111\ra)}.$ Since
$Q_{1111}=Q_{2222}=Q_{3333}=1,$
$Q_{1122}=Q_{1212}=Q_{2112}=Q_{2121}=Q_{1221}=Q_{2211}=-1,$ the rest
$Q_{ijkh}=0$ and
${|{\overrightarrow{s_i}}|}^2+{|{\overrightarrow{t_i}}|}^2=1,$
${\overrightarrow{s_i}}\cdot{\overrightarrow{t_i}}=0,$ thus to
attain the maximum of $\omega$, the third components of
$\overrightarrow{s_i}$ and $\overrightarrow{t_i}$ should be zero,
and either $|{\overrightarrow{s_i}}|=|{\overrightarrow{t_i}}|$ or
one of the $|{\overrightarrow{s_i}}|$ and $|{\overrightarrow{t_i}}|$
is zero and the other one is $1$. Let $i\neq j\neq k\neq l\in
\{1,2,3,4\}$, we can obtain the following classifications:
$$
\begin{array}{l}
1.~|{\overrightarrow{s_i}}|=1~~ (0 \ or\  \frac{1}{\sqrt{2}}),\
|\overrightarrow{s_j}|=|\overrightarrow{s_k}|=|\overrightarrow{s_l}|=1~~
(0\  or\ \frac{1}{\sqrt{2}}),\\[3mm]
2.~|{\overrightarrow{s_i}}|=|\overrightarrow{s_j}|=0,\
|\overrightarrow{s_k}|=|\overrightarrow{s_l}|=\frac{1}{\sqrt{2}},\\[3mm]
3.~|{\overrightarrow{s_i}}|=|\overrightarrow{s_j}|=1,\
|\overrightarrow{s_k}|=|\overrightarrow{s_l}|=0~~ (\ or \
\frac{1}{\sqrt{2}}),\\[3mm]
4.~|{\overrightarrow{s_i}}|=|\overrightarrow{s_j}|=1,\
|\overrightarrow{s_k}|=0,\
|\overrightarrow{s_l}|=\frac{1}{\sqrt{2}},\\[3mm]
5.~|{\overrightarrow{s_i}}|=|\overrightarrow{s_j}|=0,\
|\overrightarrow{s_k}|=1,\
|\overrightarrow{s_l}|=\frac{1}{\sqrt{2}},\\[3mm]
6.~|{\overrightarrow{s_i}}|=|\overrightarrow{s_j}|=\frac{1}{\sqrt{2}},\
|\overrightarrow{s_k}|=0,\   \ |\overrightarrow{s_l}|=1.
\end{array}\eqno{(9)}
$$

For the case $|{\overrightarrow{s_1}}|=|\overrightarrow{s_2}|=|\overrightarrow{s_3}|=|\overrightarrow{s_4}|=1$, we have
$$\omega=4{(\overrightarrow{s_1}\otimes\overrightarrow{s_2}\otimes\overrightarrow{s_3}\otimes\overrightarrow{s_4}\cdot\overrightarrow{Q})}^2=
4{{\langle\psi|{C_1C_2C_3C_4}|\psi\rangle}}^2\leq4.$$

For the case $|{\overrightarrow{s_1}}|=|{\overrightarrow{s_2}}|=|{\overrightarrow{s_3}}|=1$
 and $|{\overrightarrow{s_4}}|=\frac{1}{\sqrt{2}}$, we get
$$
\begin{array}{lll}
\omega&=&3{({\overrightarrow{s_1}\otimes\overrightarrow{s_2}\otimes\overrightarrow{s_3}\otimes
(\overrightarrow{t_4}+\overrightarrow{s_4})\cdot\overrightarrow{Q}}})^2+
{(\overrightarrow{s_1}\otimes\overrightarrow{s_2}\otimes\overrightarrow{s_3}\otimes
\overrightarrow{s_4}\cdot\overrightarrow{Q})}^2\\[3mm] &=&3{\langle\psi|
C_1C_2C_3(D_4+C_4)|\psi\rangle}^2+{\langle\psi|
C_1C_2C_3C_4|\psi\rangle}^2\leq4.\end{array}
$$

For the case $|{\overrightarrow{s_1}}|=1$  and
 $|{\overrightarrow{s_2}}|=|{\overrightarrow{s_3}}|=|{\overrightarrow{s_4}}|=\frac{1}{\sqrt{2}}$,
using the orthogonal relation of ${\overrightarrow{s_i}}$ and
${\overrightarrow{t_i}},$ we can express ${\overrightarrow{s_i}}$ and
${\overrightarrow{t_i}}$ as
$$
\begin{array}{l}
\overrightarrow{s_1}=(\sin{\theta_1}\cos{\varphi_1},\sin{\theta_1}\sin{\varphi_1},\cos{\theta_1}),\\[3mm]
\overrightarrow{s_i}=\frac{1}{\sqrt{2}}(\sin{\theta_i}\cos{\varphi_i},\sin{\theta_i}\sin{\varphi_i},\cos{\theta_i}),\\[3mm]
\overrightarrow{t_i}=\frac{1}{\sqrt{2}}(\cos{\theta_i}\cos{\varphi_i},\cos{\theta_i}\sin{\varphi_i},-\sin{\theta_i}),
\end{array}
$$
where $i=2,3,4.$
After straightforward calculations, we have
\begin{eqnarray}
  \omega &=& {[\frac{1}{2{\sqrt{2}}}\sin{(\theta_1+\varphi_1+\theta_2+\varphi_2+\theta_3+\varphi_3+\theta_4+\varphi_4)}
  }\nonumber\\
&&{-\cos{(\theta_1+\varphi_1+\theta_2+
\varphi_2+\theta_3+\varphi_3+\theta_4+\varphi_4)}]}\nonumber \\
&&+\frac{4}{2{\sqrt{2}}}\cos{\theta_1}\cos{\theta_4}{[\cos{(\theta_2+\theta_3)}-\sin{(\theta_2
+\theta_3)}]}^2\leq 4. \nonumber
\end{eqnarray}

For the case $|{\overrightarrow{s_1}}|=|{\overrightarrow{s_2}}|=|{\overrightarrow{s_3}}|=|{\overrightarrow{s_4}}|=\frac{1}{\sqrt{2}}$,
one can set
$$
\begin{array}{l}\overrightarrow{s_i}=\frac{1}{\sqrt{2}}{(\sin{\theta_i}\cos{\varphi_i},\sin{\theta_i}\sin{\varphi_i},
\cos{\theta_i})},\\[3mm]
\overrightarrow{t_i}=\frac{1}{\sqrt{2}}{(\cos{\theta_i}\cos{\varphi_i},\cos{\theta_i}\sin{\varphi_i},-\sin{\theta_i})}
\end{array}
$$
for $i=1,2,3,4$, and obtain $\omega\leq4$. For other cases inequality (\ref{t4}) can be similarly proved.
\quad $\Box$

%\vspace{2cm}
\begin{figure}[!htp]
\centerline{\includegraphics[width=0.3\textwidth]{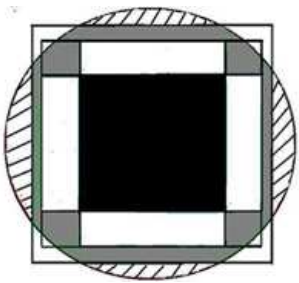}}
\caption{Projection of the state space onto the plane constituted by
$\langle D_4^{(1)}\rangle$ and $\langle D_4^{(3)}\rangle$. The fully
separable states are in the black region. The tri-separable states
$\rho_{12-3-4}$ and $\rho_{14-2-3}$ are located inside the white
areas respectively between black and gray areas. The bi-separable
states $\rho_{12-34}$, $\rho_{14-23}$, $\rho_{1-234}$ and
$\rho_{3-124}$ are inside the gray area. The genuine entangled
states are located in the designated slash regions.} \label{fig1}
\end{figure}

The inequalities in Theorems 1 to 4 give rise to an explicit
geometric picture. Taking the average of $D_{4}^{(1)}$,
$D_{4}^{(2)}$, $D_{4}^{(3)}$ and $D_{4}^{(4)}$ as the four
coordinates of a four dimensional space, from Theorem 4 we have that
$\langle{D_4}^{(i)}\rangle$, $i=1,...,4$, constitute a 3-dimensional
sphere. All fully separable states are confined in the center, in a
4-dimensional rectangular with size $1\times 1\times 1\times 1$, see
Fig. 1. While from the Theorem $2$, the tri-separable state, for
example, $\rho_{12-3-4}$, is in a 4-dimensional rectangular with the
size $1\times1\times3/2\times 3/2$. From the Theorem $3$, the
bi-separable state in class $i)$, for example $\rho_{12-34}$, is in
a 4-dimensional rectangular with size
$3/2\times3/2\times3/2\times3/2$. For the bi-separable state in
class $ii)$, for example $\rho_{1-234}$, we have a 4-dimensional
rectangular with size $\sqrt{3}\times \sqrt{3}\times \sqrt{3}\times
\sqrt{3}$.

\section {Conclusion and discussions}

We have investigated the classification of four-qubit entanglement
in terms of Bell inequalities that involving only two measurement
settings per observer. And a quadratic inequality of Bell operator
for four-qubit systems has been obtained.
The Bell inequalities satisfied by fully separable, bi-separable and
tri-separable states of four-qubit systems are analytically derived.
Our approach and some of the obtained formulas can be directly generalized
to multipartite qubit systems.

However, our inequalities are not both sufficient and necessary for
separability of general four-qubit states. The separability problem
in terms of Bell inequalities has been solved only for two-qubit
case, any pure entangled two-qubit states violate the CHSH
inequality \cite{7}, as well as the three-qubit case where Chen et
al. showed that all pure entangled three-qubit states violate a Bell
inequality \cite{9}. For mixed four-qubit systems, the separability
problem remains open. Endrejat et al. discussed in \cite{jendrejat}
the relations between optimization operators and combination of the
global entanglement measures. One may conjecture that to make the
inequalities sufficient conditions for separability of a four-qubit
mixed state, in addition to our Bell operators, some new Bell
operators are needed.

\bigskip
{\noindent\bf Acknowledgement}

\noindent H. Zhao acknowledges discussions with Ming Li. This work
was supported by the National Natural Science Foundation of China
(11101017 and 11275131), Beijing Natural Science Foundation Program
and Scientific Research Key Program of Beijing Municipal Commission
of Education (KZ201210028032).


\begin{thebibliography}{99}
\bibitem{Bell64} Bell J S. On the Einstein-Podolsky-Rosen Paradox. Physics, 1964, 1:
195-200

\bibitem{MABK} Gisin N, Bechmann-Pasquinucci H. Bell inequality,
Bell states and maximally entangled states for n qubits. Phys Lett
A, 1998, 246: 1-6

\bibitem{reviews} Paterek T, Laskowski W,
\.{Z}ukowski M. On Series of Multiqubit Bell Inequalities. Mod Phys
Lett A, 2006, 21: 111-126

\bibitem{sci} Yu S X, Chen Q, Zhang C J, et al. All entangled pure
states violate a single Bell's inequality. Phys Rev Lett, 2012, 109:
120402-120406

\bibitem{sci12} Pal K F, Vertesi T. Multisetting Bell-type inequalities for
detecting genuine tripartite entanglement. Phys Rev A, 2011, 83:
062123-062129

\bibitem{csb1} Li M, Fei S M, Li-Jost X Q. Bell inequality, Separability and
entanglement distillation. Chin Sci Bull, 2011, 56: 945-954

\bibitem{csb2}  Di Y M, Liu S P, Liu D D. Entanglement for a two-parameter class
of states in a high-dimension bipartite quantum system. Sci China
Phys Mech Astron, 2010, 53: 1868-1872

\bibitem{csb3} Guo Y, Qi X F, Hou J C. Sufficient and necessary conditions of
separability for bipartite pure states in infinite-dimensional
systems. Chin Sci Bull, 2011, 56: 840-846

\bibitem{csb4} Li X K, Li J L, Liu B, et al. The parametric symmetry and
numbers of the entangled class of $2\times M \times N$ system. Sci
China Phys Mech Astron, 2011, 54: 1471-1475

\bibitem{csb5} Wang Y Z, Hou J C, Guo Y. An entanglement criterion for states
in infinite-dimensional multipartite quantum systems. Chin Sci Bull,
2012, 57: 1643-1647

\bibitem{sci13} He Q Y, Cavalcanti E G, Reid M D, et al. Bell inequalities
for Continuous-Variable Measurements. Phys Rev A, 2010, 81:
062106-062120

\bibitem{Chen2006} Chen K, Albeverio S, Fei S M. Two-setting Bell inequalities for many
qubits. Phys Rev A, 2006, 74: 050101-050104(R)

\bibitem{lkch00} Sun B Z, Fei S M. Bell inequalities classifying biseparable three-qubit
states. Phys Rev A, 2006, 74: 032335-032338

\bibitem{cfwu} Wu C F, Yeo Y, Kwek L C, et al. Quantum nonlocality of four-qubit
entangled states. Phys Rev A, 2007, 75: 032332-032337

\bibitem{yyeo} Yeo Y, Chua W K. Teleportation and Dense Coding with Genuine Multipartite
Entanglement. Phys Rev Lett, 2006, 96: 060502-060505

\bibitem{yxxia} Yu S X, Pan J W, Chen Z B, et al. Comprehensive test of entanglement
for two-level systems via the indeterminacy relationship. Phys Rev
Lett, 2003, 91: 217903-217906

\bibitem{knagata} Nagata K, Koashi M, Imoto N. Configuration of Separability and
Tests for Multipartite Entanglement in Bell-Type Experiments. Phys
Rev Lett, 2002, 89: 260401-260404

\bibitem{sci2} Brunner N, Sharam J, Vertesi T. Testing the Structure of
Multipartite Entanglement with Bell Inequalities. Phys Rev Lett,
2012, 108: 110501-110505

\bibitem{sci21} Chen J L, Deng D L, Su H Y, et al. Detecting Full
N-Particle Entanglement in Arbitrarily High-Dimensional Systems with
Bell-Type Inequality. Phys Rev A, 2011, 83: 022316-022321

\bibitem {nha} Nha H, Zubairy M S. Uncertainty Inequalities as Entanglement Criteria for
Negative Partial-Transpose States. Phys Rev Lett, 2008, 101:
130402-130405

\bibitem{SG2001} Scarani V, Gisin N. Spectral decomposition of Bell's operators for
qubits. J Phys A, 2001, 34: 6043-6053

\bibitem{ZBLW2002} \.{Z}ukowski M,
Brukner \v{C}. Laskowski W, et al. Do all pure entangled states
violate Bells inequalities for correlation functions? Phys Rev Lett,
2002, 88: 210402-210405

\bibitem{wdur} D$\ddot{u}$r W, Vidal G, Cirac J I. Three qubits can be entangled
in two inequivalent ways. Phys Rev A, 2000, 62: 062314-062325

\bibitem{linden} Linden N, Popescu S, Sudbery A. Nonlocal Parameters for Multiparticle
Density Matrices. Phys Rev Lett, 1999, 83: 243-247

\bibitem{rudolph} Carteret H A, Higuchi A, Sudbery A. Multipartite generalisation of the
Schmidt decomposition. J Math Phys, 2000, 41: 7932-7939

\bibitem{7} Gisin N. Bell's inequality holds for all non-product states. Phys Lett A, 1991, 154: 201-202

\bibitem{9} Chen J L, Wu C F, Kwek L C, et al. Gisin's Theorem for three qubits.
Phys Rev Lett, 2004, 93: 140407-140410

\bibitem{jendrejat} Endrejat J, B$\ddot{u}$ttner H. Characterization of entanglement of more than two qubits with Bell
inequalities and global entanglement. Phys Rev A, 2005, 71:
012305-012313
\end{thebibliography}
\end{document}